\documentclass[preprint, superscriptaddress,preprintnumbers,amsmath,amssymb,pra]{revtex4}
\usepackage{graphicx}
\usepackage{amsmath}

\begin{document}

\thispagestyle{empty}

\title{Influence of chemical potential on the Casimir-Polder interaction
between an atom and gapped graphene or graphene-coated substrate}

\author{C.~Henkel}
\affiliation{Institute of Physics and Astronomy, University of Potsdam,
Karl-Liebknecht-Stra\ss{}e 24/25, 14476, Potsdam, Germany}

\author{
G.~L.~Klimchitskaya}
\affiliation{Central Astronomical Observatory at Pulkovo of the
Russian Academy of Sciences, Saint Petersburg,
196140, Russia}
\affiliation{Institute of Physics, Nanotechnology and
Telecommunications, Peter the Great Saint Petersburg
Polytechnic University, Saint Petersburg, 195251, Russia}

\author{
V.~M.~Mostepanenko}
\affiliation{Central Astronomical Observatory at Pulkovo of the
Russian Academy of Sciences, Saint Petersburg,
196140, Russia}
\affiliation{Institute of Physics, Nanotechnology and
Telecommunications, Peter the Great Saint Petersburg
Polytechnic University, Saint Petersburg, 195251, Russia}
\affiliation{Kazan Federal University, Kazan, 420008, Russia}

\begin{abstract}
We present a formalism based on first principles of
quantum electrodynamics at nonzero temperature which permits to
calculate the Casimir-Polder interaction between an atom
and a graphene sheet
with arbitrary mass gap and chemical potential, including
graphene-coated substrates.
The free energy and
force of the Casimir-Polder interaction are expressed via the
polarization tensor of graphene in (2+1)-dimensional space-time
in the framework of the Dirac model. The obtained expressions
are used to investigate the influence of the chemical potential
of graphene on the Casimir-Polder interaction.
Computations are performed for an atom of metastable helium
interacting with either a free-standing graphene sheet or a
graphene-coated substrate made of amorphous silica. It is
shown that the impacts of the nonzero chemical potential and
the mass gap on the Casimir-Polder interaction are in opposite
directions by increasing and decreasing the magnitudes of the
free energy and force, respectively. It turns out, however,
that the temperature-dependent part of the Casimir-Polder interaction
is decreased by a nonzero chemical potential, whereas the
mass gap increases it compared to the case of undoped,
gapless graphene. The physical explanation for these effects
is provided. Numerical computations of the Casimir-Polder
interaction are performed at various temperatures and
atom-graphene separations.

\end{abstract}

\maketitle

\section{Introduction}
\newcommand{\hem}{{He${}^{\ast}${\ }}}

With the advent of graphene, which is a two-dimensional sheet
of carbon atoms packed in a hexagonal lattice, it has found
widespread application in both fundamental and applied
physics \cite{1,2}. One of the subjects of much recent
attention is the interaction of graphene with the zero-point
and thermal fluctuations of the electromagnetic field giving
rise to the van der Waals (Casimir) and Casimir-Polder
forces \cite{3,4}.
These forces act between two graphene sheets and an atom
and a graphene sheet (or graphene-coated substrate), respectively.
Given that the optical properties of
graphene can be modified by doping,
it may be possible to tune both the van der Waals (Casimir)
and Casimir-Polder interactions.
The van der Waals and Casimir interactions
between two graphene sheets, a graphene sheet and a
3D-material plate, and graphene-coated substrates have been
investigated in the framework of the Dirac model. This model assumes
that at low energies the graphene quasiparticles obey a
linear dispersion relation but move with the Fermi velocity
$v_F \approx c/300$ rather than with the speed of light
\cite{1,2,5}. A lot of calculations were performed using the
density-density correlation functions, the Kubo formalism,
and some special models for the dielectric permittivity
(conductivity) of graphene
\cite{6,7,8,9,10,11,12,13,14,15,16,17,18,18a}.
The same methods have been used to calculate
the Casimir-Polder force between different atoms and
graphene sheet under various conditions
\cite{19,20,21,22,23,23a}.
Specifically, in Ref.~\cite{23a} the Casimir-Polder interaction
between an atom and a substrate coated with a charge layer
was considered. This layer characterized by a nonlocal dielectric
response can be used as a simplified model of a graphene sheet.

In the framework of the Lifshitz theory of dispersion
forces \cite{2,24}, the Casimir and Casimir-Polder
interactions can be expressed in terms
of the reflection coefficients for electromagnetic
fluctuations. Within the Dirac model, the reflection
coefficients of graphene are expressed
via a polarization tensor in (2+1)-dimensional
space-time \cite{25,26}. This model allows for both
zero and nonzero quasiparticle mass $m$. The latter may arise
due to electron-electron interactions, impurities and the
presence of a substrate.
The polarization tensor of Refs.~\cite{25,26}
has been used to calculate the Casimir force
in many physical systems incorporating graphene sheets
for any mass gap $\Delta = 2 m c^{2}$ and at any temperature
\cite{25,26,27,28,30,31}, but it is restricted to the
case of undoped graphene
(chemical potential $\mu = 0$).
The computational results were found \cite{32,33} to be
in a very good agreement with the experimental data of
the work on measuring the gradient of the Casimir force
between an Au-coated sphere and a graphene-coated
substrate \cite{34}. The same polarization tensor was
applied to investigate the Casimir-Polder interaction
of different atoms with gapped graphene \cite{35,36,37,38}
and with graphene-coated plates made of different
materials \cite{39}. The classical limit of the
Casimir-Polder interaction with graphene systems has also
been considered \cite{40}.

The polarization tensor of graphene of Ref.~\cite{26} is
restricted to the purely imaginary Matsubara frequencies.
Another representation that provides the analytic continuation
to the entire complex frequency plane was derived in
Ref.~\cite{41}.
This
representation has been used to investigate
the thermal Casimir force in graphene systems
\cite{42,43,44,45}, the electrical conductivity of both
gapless and gapped graphene \cite{46,47}, and the
reflectivity properties of graphene and graphene-coated
substrates \cite{41,48,49,50}.

Real graphene samples are always doped and can be
characterized by a nonzero chemical potential $\mu$
\cite{5}. Because of this, it is desirable to describe
the Casimir and Casimir-Polder forces in graphene systems
with account of both parameters $\Delta$ and $\mu$.
In Ref.~\cite{52} the polarization tensor of graphene
found in Ref.~\cite{41} was generalized for the case of
doped graphene with nonzero chemical potential. According
to Ref.~\cite{52}, the thermal Casimir force between the
doped but gapless graphene sheet and an ideal-metal plane
can be enhanced up to 60\% in comparison to the case of
undoped graphene. The detailed investigation of the
thermal Casimir force in graphene systems with nonzero
mass gap and chemical potential demonstrated that these
parameters act in the opposite directions by decreasing
and increasing the force magnitude, respectively
\cite{53}. However, the role of the chemical potential
in the Casimir-Polder interaction between an atom and
a graphene sheet or a graphene-coated substrate remained
unexplored.

In this paper, we investigate the Casimir-Polder
interaction between an atom and a graphene sheet or a
graphene-coated substrate in thermal equilibrium with
the environment. Graphene is described in the framework
of the Dirac model by the polarization tensor taking into
account the mass gap and chemical
potential at any temperature. In doing so, we consider
not too small atom-graphene separations in order to
remain in the application region of the Dirac model,
where the dispersion relation for graphene quasiparticles
remains linear (this holds at energies below 1--2\,eV
\cite{5,26,53a}).
The material of a substrate is described
by a local isotropic dielectric function.

We present the expressions for the Casimir-Polder
free energy and force based on first principles of
quantum electrodynamics at nonzero temperature. The
expressions are used to compute the
Casimir-Polder interaction between an atom of
metastable helium \hem and a graphene sheet
characterized by various values of the mass gap and
chemical potential. This atom has a relatively large polarizability
and has been used in quantum reflection experiments that
are sensitive to the atom-surface interaction
\cite{53b,53c}.
Similar computations are performed
for an atom of \hem interacting with a
graphene-coated SiO${}_2$ substrate. All computations are made
at room temperature and at liquid nitrogen temperature.
It is shown that with increasing mass gap or chemical
potential the magnitudes of both the Casimir-Polder
free energy and force decrease or increase, respectively.
Thus, the impacts of both parameters on the
Casimir-Polder interaction are in the opposite
directions and partially compensate each other. This
result is important from the experimental point of view.
From the theoretical viewpoint, the temperature dependence
of the Casimir-Polder force has attracted much interest, but is
usually manifests itself at relatively large distance. For
graphene, the situation is more favorable because the
thermal regime is reached at shorter distances (well below
one micrometer). We find that a larger chemical
potential suppresses the role of thermal correction at all
separations. By contrast, for a
larger mass gap the thermal effect is
larger. For a graphene-coated substrate,
the Casimir-Polder interaction is stronger than for a bare substrate,
in particular if the latter is dielectric,
but possesses similar physical
properties compared to a free-standing graphene sheet.

The paper is organized as follows. In Sec. II, the
exact formalism is presented including the analytic
expressions for the Casimir-Polder free energy and force
in terms of the polarization tensor of graphene with
nonzero $\Delta$ and $\mu$. Section III contains the
results of numerical computations of the Casimir-Polder
free energy and force between an atom of \hem
and a free-standing graphene sheet. In Sec. IV, similar
results for an atom of \hem interacting with
a graphene-coated SiO${}_2$ substrate are presented.
Section V contains our conclusions and a discussion.

\section{Exact formalism in the framework of Dirac model}

\newcommand{\ve}{\varepsilon}
\newcommand{\fe}{{\cal F}}
\newcommand{\tp}{{\tilde{\Pi}}}
\newcommand{\vf}{{{\tilde{v}}_F}}

We consider an atom characterized by the frequency-dependent
isotropic  electric
dipole polarizability
$\alpha(\omega)$
at a distance $a$ from a graphene sheet deposited on a thick
material substrate (semispace) described by the frequency-dependent dielectric
permittivity $\ve(\omega)$. Graphene is characterized by the mass-gap parameter
$\Delta=2mc^2$, where $m$ is the mass of quasiparticles, and chemical potential
$\mu$. The considered system is assumed to be in thermal equilibrium with the
environment at temperature $T$. The Casimir-Polder free energy of an atom interacting
with a graphene-coated substrate is given by the Lifshitz formula \cite{4,24} which we
present in terms of the dimensionless variables
\begin{eqnarray}
&&
{\cal F}(a,T)=-k_BT\sum_{l=0}^{\infty}
{\vphantom{\sum}}^{\prime}
\alpha(i\zeta_l\omega_c)
\mathop{\rm tr}{G}( a, {i} \zeta_l\omega_c ),
\label{eq1}\\
&&
\mathop{\rm tr}{G}( a, {i} \zeta_l\omega_c )
=
\frac{1}{8a^3}
\int_{\zeta_l}^{\infty}\!\!\!dye^{-y}
\left\{2y^2R_{\rm TM}(i\zeta_l,y)-\zeta_l^2\left[
R_{\rm TM}(i\zeta_l,y)+R_{\rm TE}(i\zeta_l,y)\right]
\right\}.
\nonumber
\end{eqnarray}
\noindent
Here, $k_B$ is the Boltzmann constant,
the prime on the summation sign indicates
that the term with $l=0$ is divided by two, and
the dimensionless Matsubara frequencies  are
$\zeta_l=\xi_l/\omega_c$,
where $\xi_l=2\pi k_BTl/\hbar$ with
$l=0,\,1,\,2,\,\ldots$ are the dimensional Matsubara
frequencies and  $\omega_c=c/(2a)$.
The electromagnetic Green tensor ${G}$
(the free-space contribution to ${G}$ is subtracted)
describes how
the field emitted by the atomic dipole is reflected
by the surface, as encoded in the reflection amplitudes
$R_{\rm TM}$ and $R_{\rm TE}$
for two independent polarizations, transverse
magnetic (TM) and transverse electric (TE).
Note that the dimensionless integration variable $y$ is connected
with the magnitude of the projection of the wave vector on the plane of
graphene, $k_{\bot}$, by $y=2aq_l$ where $q_l^2=k_{\bot}^2+\xi_l^2/c^2$.

The remaining undefined quantities in Eq.~(\ref{eq1}) are the reflection coefficients. They are expressed through the dielectric
permittivity of a substrate material
$\ve_l\equiv\ve(i\xi_l)=\ve(i\zeta_l\omega_c)$
and the dimensionless polarization tensor of graphene $\tp_{\beta\gamma}$ with
$\beta,\,\gamma=0,\,1,\,2$ connected with the dimensional tensor $\Pi_{\beta\gamma}$
by
\begin{equation}
\tp_{\beta\gamma,l}\equiv\tp_{\beta\gamma}(i\zeta_l,y)=
\frac{2a}{\hbar}\Pi_{\beta\gamma}(i\xi_l,k_{\bot}).
\label{eq2}
\end{equation}
\noindent
As the two independent components of $\Pi_{\beta\gamma}$, it is customary to
choose $\Pi_{00}$ and ${\rm tr}\Pi$, where ${\rm tr}\Pi=\Pi_{\beta}^{\,\beta}$
is the trace of the polarization tensor.
For our purposes, however, it is more convenient to consider, instead of
${\rm tr}\Pi$, the following combination
\begin{equation}
\Pi_l=k_{\bot}^2{\rm tr}\Pi_{l}-q_l^2\Pi_{00,l}.
\label{eq3}
\end{equation}
\noindent
In terms of the dimensionless quantities, Eq.~(\ref{eq3}) reduces to
\begin{equation}
\tp_l=\frac{(2a)^3}{\hbar}\tilde{\Pi}_l=
(y^2-\zeta_l^2){\rm tr}\tilde{\Pi}_{l}-y^2\tp_{00,l}.
\label{eq4}
\end{equation}

The polarization tensor is directly connected with the nonlocal
dielectric permittivities along the graphene surface
(the longitudinal one) and
perpendicular to it (the transverse one) \cite{31}
\begin{eqnarray}
&&
\varepsilon_{\rm long}(i\xi_l,k_{\bot})=1+
\frac{1}{2\hbar k_{\bot}}\Pi_{00}(i\xi_l,k_{\bot}),
\nonumber\\[1mm]
&&
\varepsilon_{\rm tr}(i\xi_l,k_{\bot})=1+
\frac{c^2}{2\hbar k_{\bot}\xi_l^2}\Pi(i\xi_l,k_{\bot}).
\label{eq6a}
\end{eqnarray}

Taking into account the importance of the reflection coefficients
in this formalism, we present them first in terms of
dimensional variables
 \cite{32,53}
\begin{eqnarray}
&&
R_{\rm TM}(i\xi_l,k_{\bot})=
\frac{
\varepsilon_lq_l - k_l +
\frac{ q_lk_l\Pi_{00,l} }{ \hbar k_{\bot}^2 }
}{
\varepsilon_lq_l + k_l +
\frac{ q_lk_l\Pi_{00,l} }{ \hbar k_{\bot}^2 }
},
\nonumber \\
&&
R_{\rm TE}(i\xi_l,k_{\bot})=
\frac{
q_l - k_l - \frac{ \Pi_l }{ \hbar k_{\bot}^2 } }{
q_l + k_l + \frac{ \Pi_l }{ \hbar k_{\bot}^2 }
},
\label{eq4a}
\end{eqnarray}
\noindent
where
$k_l=\sqrt{k_{\bot}^2+\varepsilon_l\xi_l^2/c^2}$.

For the case of an atom interacting with a free-standing graphene sheet,
we put $\ve_l=1$ and Eq.~(\ref{eq4a}) simplifies to \cite{26,56a}
\begin{eqnarray}
&&
R_{\rm TM}(i\xi_l,k_{\bot})=
\frac{q_l\Pi_{00,l}}{q_l\Pi_{00,l}+2\hbar k_{\bot}^2},
\nonumber \\
&&
R_{\rm TE}(i\xi_l,k_{\bot})=-
\frac{\Pi_l}{\Pi_l+2\hbar k_{\bot}^2q_l}.
\label{eq4b}
\end{eqnarray}
\noindent
If there is no graphene coating, $\Pi_{00,l}=\Pi_l=0$,
and Eq.~(\ref{eq4a}) returns us back to the standard (Fresnel)
reflection coefficients.

In terms of dimensionless variables introduced above,
Eq.~(\ref{eq4a}) takes the form
\begin{eqnarray}
&&
R_{\rm TM}(i\zeta_l,y)=\frac{\varepsilon_l y(y^2-\zeta_l^2)+
\sqrt{y^2+(\ve_l-1)\zeta_l^2}\left[y\tilde{\Pi}_{00,l}-
(y^2-\zeta_l^2)\right]}{\varepsilon_l y(y^2-\zeta_l^2)+
\sqrt{y^2+(\ve_l-1)\zeta_l^2}\left[y\tilde{\Pi}_{00,l}+
(y^2-\zeta_l^2)\right]},
\nonumber \\
&&
R_{\rm TE}(i\zeta_l,y)=
\frac{(y^2-\zeta_l^2)[y-\sqrt{y^2+(\ve_l-1)\zeta_l^2}]-
\tilde{\Pi}_l}{(y^2-\zeta_l^2)[y+\sqrt{y^2+(\ve_l-1)\zeta_l^2}]+
\tilde{\Pi}_l}
\label{eq5}
\end{eqnarray}
and Eq.~(\ref{eq4b}) can be written as
\begin{eqnarray}
&&
R_{\rm TM}(i\zeta_l,y)=
\frac{y\tilde{\Pi}_{00,l}}{y\tilde{\Pi}_{00,l}+2(y^2-\zeta_l^2)},
\nonumber \\
&&
R_{\rm TE}(i\zeta_l,y)=-
\frac{\tilde{\Pi}_{l}}{\tilde{\Pi}_{l}+2y(y^2-\zeta_l^2)}.
\label{eq6}
\end{eqnarray}

The explicit expressions for $\tp_{00,l}$ and $\tp_l$ for graphene with nonzero
chemical potential were found in Ref.~\cite{52} and used in Ref.~\cite{53}
to investigate the joint action of $\Delta$ and $\mu$ on the thermal Casimir
force. By using the dimensionless variables $y$ and $\zeta_l$, we represent
the respective equations of Ref.~\cite{53} in a more simple form.
At first it is convenient to write the quantities $\tp_{00,l}$ and $\tp_l$ as
sums of two contributions
\begin{eqnarray}
&&
\tp_{00,l}(y,T,m,\mu)=\tp_{00,l}^{(0)}(y,m)
+\tp_{00,l}^{(1)}(y,T,m,\mu),
\nonumber \\
&&
\tp_l(y,T,m,\mu)=\tp_l^{(0)}(y,m)
+\tp_l^{(1)}(y,T,m,\mu),
\label{eq7}
\end{eqnarray}
\noindent
where the first terms on the right-hand side refer to undoped graphene
with $\mu=0$ at zero temperature, whereas the second ones account for the
thermal effect and for the dependence on $\mu$. Note that $\tp_{00,l}^{(1)}$
and $\tp_{l}^{(1)}$ may remain different from zero even in the limiting case
$T\to 0$ and, thus, have no meaning of thermal corrections.

The explicit form for $\tp_{00,l}^{(0)}$
and $\tp_{l}^{(0)}$ is the following \cite{25,26,53}
\begin{eqnarray}
&&
\tp_{00,l}^{(0)}(y,m)=\alpha \frac{y^2-\zeta_l^2}{p_l}\,\Psi(D_l),
\nonumber \\
&&
\tp_{l}^{(0)}(y,m)=\alpha (y^2-\zeta_l^2){p_l}\,\Psi(D_l),
\label{eq8}
\end{eqnarray}
\noindent
where $\alpha=e^2/(\hbar c)$ is the fine structure constant,
$D_l=4mca/(\hbar p_l)$,
\begin{eqnarray}
&&
\Psi(x)=2\left[x+(1-x^2)\arctan\frac{1}{x}\right],
\nonumber \\
&&
p_l\equiv p_l(y)=\sqrt{\vf^2y^2+(1-\vf^2)\zeta_l^2},
\label{eq9}
\end{eqnarray}
\noindent
and the dimensionless Fermi velocity is $\vf=v_F/c\approx 1/300$.

The explicit expressions for the second terms on the right-hand side of
Eq.~(\ref{eq7}) were derived in Ref.~\cite{52} (see also Ref.~\cite{53} for
an equivalent representation). Using the dimensionless variables,
they can be written as
\begin{eqnarray}
&&
\tp_{00,l}^{(1)}(y,T,m,\mu)=\frac{4\alpha p_l}{\vf^2}
\int_{D_l}^{\infty}du \,w_l(u,y,T,\mu)
\nonumber \\
&&\times\left\{1-{\rm Re}
\frac{p_l-p_lu^2+2i\zeta_lu}{[p_l^2-p_l^2u^2+\vf^2(y^2-\zeta_l^2)D_l^2
+2i\zeta_lp_lu]^{1/2}}\right\},
\nonumber \\
&&
\tp_{l}^{(1)}(y,T,m,\mu)=-\frac{4\alpha p_l}{\vf^2}
\int_{D_l}^{\infty}du\, w_l(u,y,T,\mu)
\label{eq10} \\
&&\times\left\{\zeta_l^2-p_l{\rm Re}
\frac{\zeta_l^2-p_l^2u^2+\vf^2(y^2-\zeta_l^2)D_l^2
+2i\zeta_lp_lu}{[p_l^2-p_l^2u^2+\vf^2(y^2-\zeta_l^2)D_l^2
+2i\zeta_lp_lu]^{1/2}}\right\}.
\nonumber
\end{eqnarray}
\noindent
Here, we have used the notation
\begin{eqnarray}
&&
w_l(u,y,T,\mu)=\frac{1}{e^{B_lu+\frac{\mu}{k_BT}}+1}+
\frac{1}{e^{B_lu-\frac{\mu}{k_BT}}+1},
\nonumber \\
&&
B_l\equiv B_l(y,T)=\frac{\hbar c}{4ak_BT}p_l(y),
\label{eq11}
\end{eqnarray}
\noindent
where $p_l$ is defined in Eq.~(\ref{eq9}).

As is seen in Eq.~(\ref{eq10}), the expressions for $\tp_{00,l}^{(1)}$
and $\tp_l^{(1)}$ with $l\geq 1$ are much more complicated than with
$l=0$, and it is convenient to deal with them separately.
Using Eqs.~(\ref{eq7}), (\ref{eq8}), and (\ref{eq10}), we first present
the total quantities $\tp_{00,l}$ and $\tp_l$ at $l=0$
\begin{eqnarray}
&&
\tp_{00,0}(y,T,m,\mu)=\frac{\alpha y}{\vf}\Psi(D_0)+
\frac{16\alpha}{\vf^2}\frac{ak_BT}{\hbar c}
\nonumber \\
&&
\times\ln\left[\left(e^{\frac{\mu}{k_BT}}+e^{-\frac{mc^2}{k_BT}}\right)
\left(e^{-\frac{\mu}{k_BT}}+e^{-\frac{mc^2}{k_BT}}\right)\right]
\nonumber\\
&&
-\frac{4\alpha y}{\vf}\int_{D_0}^{\sqrt{1+D_0^2}}
\!\!du\, w_0(u,y,T,\mu)\frac{1-u^2}{(1-u^2+D_0^2)^{1/2}},
\nonumber\\
&&
\tp_{0}(y,T,m,\mu)={\alpha{\vf} y^3}\Psi(D_0)
\label{eq12} \\
&&
+{4\alpha{\vf} y^3}\int_{D_0}^{\sqrt{1+D_0^2}}
\!\!du\, w_0(u,y,T,\mu)\frac{-u^2+D_0^2}{(1-u^2+D_0^2)^{1/2}},
\nonumber
\end{eqnarray}
\noindent
where
\begin{equation}
D_0=\frac{2 a \Delta}{\hbar v_F y}, \quad
B_0=\frac{\hbar}{4a} \frac{v_F y}{k_BT}.
\label{eq13}
\end{equation}

For $l\geq 1$ one can use much simpler approximate expressions for
$\tp_{00,l}^{(1)}$ and $\tp_{l}^{(1)}$ than those presented in Eq.~(\ref{eq10})
with no loss in accuracy. It was shown that for all $l\geq 1$ under the
condition $\zeta_1\gg\vf$
[which is equivalent to $k_BT\gg \hbar v_F/(4\pi a)$]
Eqs.~(\ref{eq7}), (\ref{eq8}) and (\ref{eq10})
lead to \cite{42,53}
\begin{eqnarray}
&&
\tp_{00,l}(y,T,m,\mu)\approx \frac{\alpha(y^2-\zeta_l^2)}{\zeta_l}
\nonumber \\
&&~~~~~~~~
\times
\left[\Psi\left(\frac{4mca}{\hbar\zeta_l}\right)+
\tilde{Y}_l(y,T,m,\mu)\right],
\label{eq14} \\
&&
\tp_{l}(y,T,m,\mu)\approx {\alpha{\zeta_l}(y^2-\zeta_l^2)}
\nonumber \\
&&~~~~~~~~
\times
\left[\Psi\left(\frac{4mca}{\hbar\zeta_l}\right)+
\tilde{Y}_l(y,T,m,\mu)\right],
\nonumber
\end{eqnarray}
\noindent
where
\begin{equation}
\tilde{Y}_l(y,T,m,\mu)=2\int_{\frac{4mca}{\hbar\zeta_l}}^{\infty}\!\!\!\!
duw_l(u,y,T,\mu)
\frac{u^2+\left(\frac{4mca}{\hbar\zeta_l}\right)^2}{u^2+1}.
\label{eq15}
\end{equation}

Note that for $a=50\,$nm at $T=300\,$K the first dimensionless Matsubara frequency
$\zeta_1$ is larger than $\vf$ by a factor of 25 (and more for larger separations).
Because of this, the use of the approximate expression (\ref{eq14}) leads to
practically
exact Casimir-Polder free energy and force (the relative error is
less
than 0.02\% \cite{42})
if the zero-frequency contributions to
them are calculated using the exact Eq.~(\ref{eq12}).
Computations show that even at liquid nitrogen temperature ($T=77.2\,$K) for
$a>50\,$nm ($\zeta_1/\vf>6.4$) the use of  Eq.~(\ref{eq14}) for $l\geq 1$ allows
computation of the Casimir-Polder interaction accurate to a fraction of a percent.

\begin{figure}[b]
\vspace*{-4cm}
\centerline{\hspace*{2.5cm}
\includegraphics{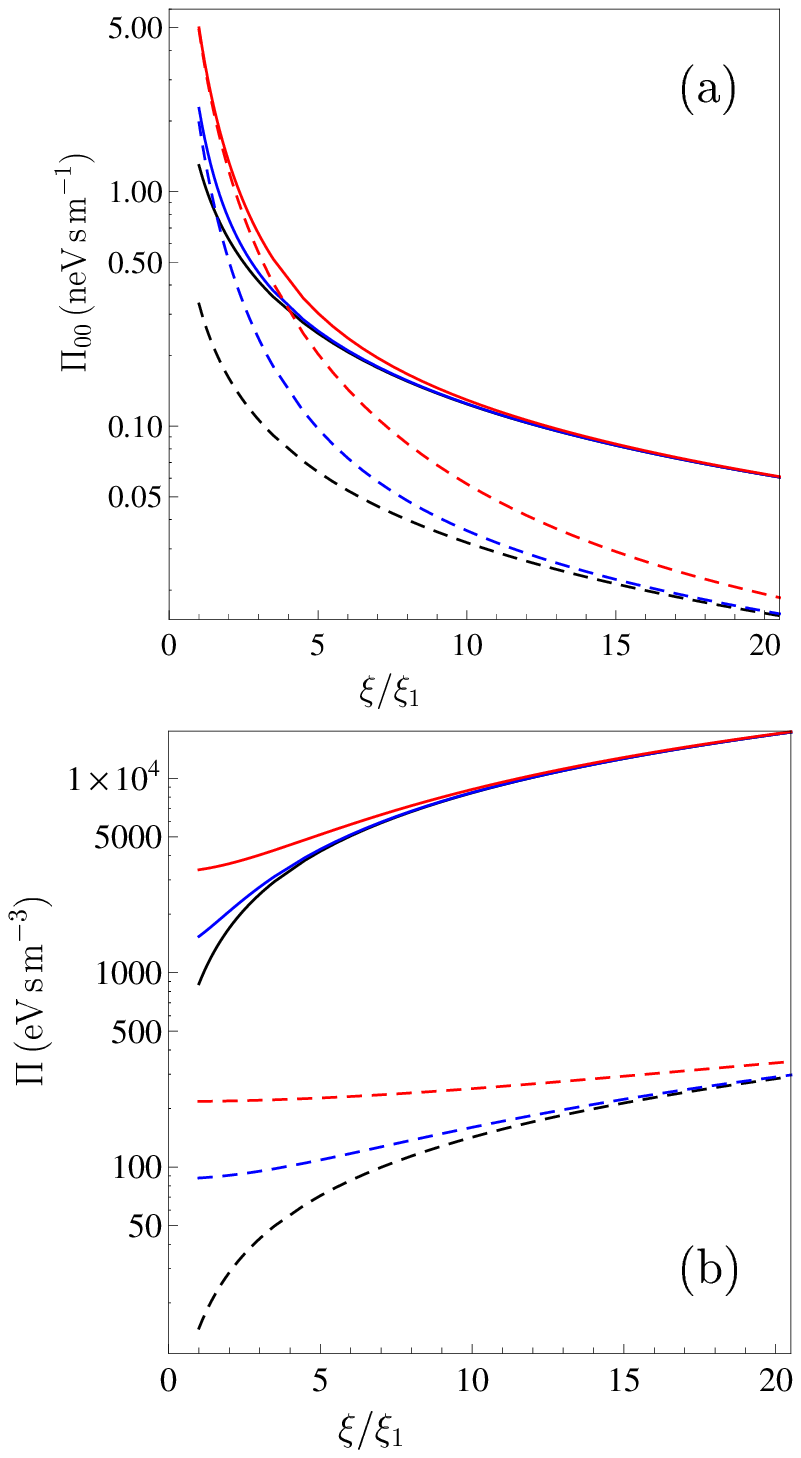}
}
\vspace*{-12cm}
\caption{\label{fg1a}
(a) The 00-component of the polarization tensor $\Pi_{00}$ and
(b) the combination of its components $\Pi$ are shown at
$T=300\,$K (the solid lines) and at $T=77\,$K (the dashed lines)
for $k_{\bot}=10\xi_1/c$ as functions of $\xi/\xi_1$ along the
imaginary frequency axis for $\xi\geq\xi_1$. The lines of each
kind from bottom to top are plotted for a gapless graphene with
$\mu=0$, 0.2, and 0.5\,eV, respectively.
Here and in all other figures, we take for the Fermi velocity
the value $v_F=c/300$.
}
\end{figure}

As an illustration, in Fig.~\ref{fg1a} we present (a) the
00-component of the polarization tensor, $\Pi_{00}(i\xi,k_{\bot})$,
and (b) the quantity $\Pi(i\xi,k_{\bot})$ defined in
Eq.~(\ref{eq3})
as functions of $\xi/\xi_1$, where $\xi$ varies along the
imaginary frequency axis
and $\xi\geq\xi_1$. Computations are performed using
Eqs.~(\ref{eq2}), (\ref{eq4}), (\ref{eq12}), and (\ref{eq14}) for
a gapless graphene ($m=0$) at $k_{\bot}=10\xi_1/c$ at room
temperature $T=300\,$K (the solid lines) and at
liquid nigrogen temperature $T=77\,$K (the dashed lines).
The lines of each kind from bottom to top are plotted for the
chemical potential $\mu=0$, 0.2, and 0.5\,eV, respectively
(see the discussion of typical values taken by the chemical
potential of graphene samples in Sec.~III). At zero Matsubara
frequency $\Pi_{00}$ takes the values 0.507, 11.0, and
$27.5\,\mu\mbox{eV\,s/m}$ ($T=77\,$K) and
1.97, 11.0, and
$27.5\,\mu\mbox{eV\,s/m}$ ($T=300\,$K) for $\mu=0$, 0.2, and
0.5\,eV, respectively. The quantity $\Pi$ at zero Matsubara
frequency takes the values 0.021, 0, and $0\,\mbox{eV\,s/m}^3$
($T=77\,$K) and 1.25, 0.0022, and $0\,\mbox{eV\,s/m}^3$
($T=300\,$K) for $\mu=0$, 0.2, and 0.5\,eV, respectively.
As is seen in Fig.~\ref{fg1a}, the quantity $\Pi_{00}$ decreases
and the quantity $\Pi$ increases monotonously with increase of
$\xi$. In all cases the magnitudes of both $\Pi_{00}$ and $\Pi$
are larger for higher temperature and larger chemical potential.
{}From Figs.~\ref{fg1a}(a) and \ref{fg1a}(b) one can conclude
that the impact of $\mu$ on the values of $\Pi_{00}$ and $\Pi$
decreases with increasing frequency.

As a result, the Casimir-Polder free energy of an atom interacting with a
free-standing graphene sheet or graphene-coated substrate can be computed by
Eqs.~(\ref{eq1}) and (\ref{eq5}) or (\ref{eq6}) where the polarization tensor
is given in Eqs.~(\ref{eq7}), (\ref{eq8}) and (\ref{eq10}) or (\ref{eq12})
and (\ref{eq14}). To calculate the respective Casimir-Polder force, one
should use the following Lifshitz formula
\cite{4}:
\begin{eqnarray}
&&
{F}(a,T)=-\frac{k_BT}{8a^4}\sum_{l=0}^{\infty}
{\vphantom{\sum}}^{\prime}
\alpha(i\zeta_l\omega_c)
\int_{\zeta_l}^{\infty}\!\!\!ydye^{-y}
\label{eq16}\\
&&~
\times\left\{2y^2R_{\rm TM}(i\zeta_l,y)-\zeta_l^2\left[
R_{\rm TM}(i\zeta_l,y)+R_{\rm TE}(i\zeta_l,y)\right]
\right\}
\nonumber
\end{eqnarray}
in place of Eq.~(\ref{eq1}).


\section{Interaction with free-standing doped graphene sheet}

\begin{figure}[b]
\vspace*{-4cm}
\centerline{\hspace*{2.5cm}
\includegraphics{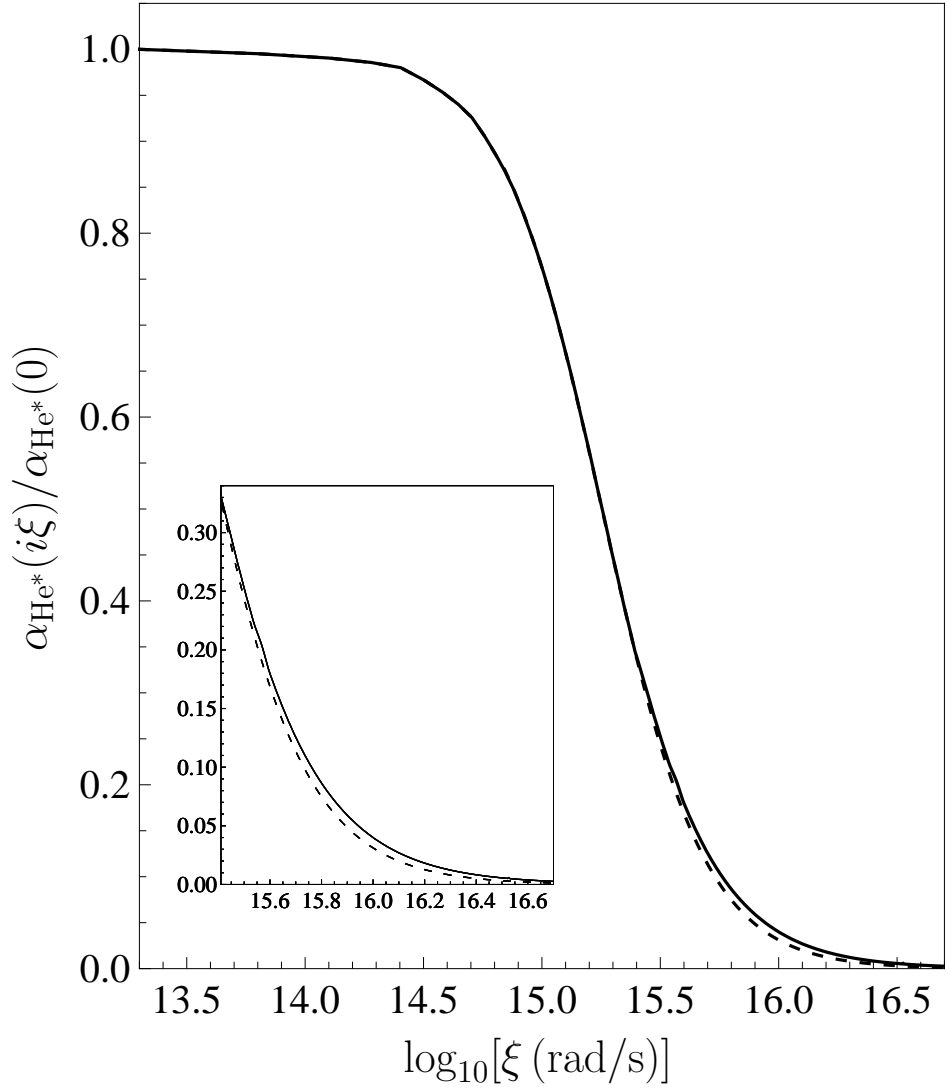}
}
\vspace*{-10cm}
\caption{\label{fg1}
The highly-accurate and single-oscillator dynamic atomic
polarizabilities of He${}^{\ast}$ normalized to their static
value are shown as functions of the imaginary frequency
by the solid and dashed lines, respectively.
The region of high frequencies is presented on an enlarged scale
in the inset.
}
\end{figure}

Here, we calculate the Casimir-Polder free energy and force for an atom of
metastable helium (He${}^{\ast}$) interacting with a free-standing graphene sheet.
All computations are performed  by using Eqs.~(\ref{eq1}) and (\ref{eq16}) where
the reflection coefficients are given by Eq.~(\ref{eq6}) and the polarization tensor by
Eqs.~(\ref{eq12}) and (\ref{eq14}),
and we vary the mass-gap parameter and the chemical potential.
These computations require data for the \hem polarizability $\alpha(i \xi)$
as a function of imaginary frequency. Below we use the highly accurate polarizability of Refs.\cite{54,55}
which has a relative error of order of $10^{-6}$. It is shown in Fig.~\ref{fg1}
by the solid line as a function of frequency,
normalized to its static value
$\alpha(0) = 46.7727\,\mbox{\AA}^3$ \cite{56}.
Note that the dynamic
polarizability is often represented using the single-oscillator model
\cite{55,56,57,58}
\begin{equation}
\alpha(i\xi)=\frac{\alpha(0)}{1+\frac{\xi^2}{\omega_0^2}},
\label{eq17}
\end{equation}
\noindent
where for \hem one has $\omega_0=1.793\times 10^{15}\,$rad/s \cite{56}.
In Fig.~\ref{fg1}, the latter one is shown by the dashed line,
and it is seen that the major deviations between the
highly accurate and the single-oscillator data for the dynamic polarizability
occur in the region of high frequencies (see inset at an
enlarged scale).
This corresponds to the atom--graphene separations of order
of 10\,nm or shorter.

\begin{figure}[b]
\vspace*{0cm}
\centerline{\hspace*{2.5cm}
\includegraphics{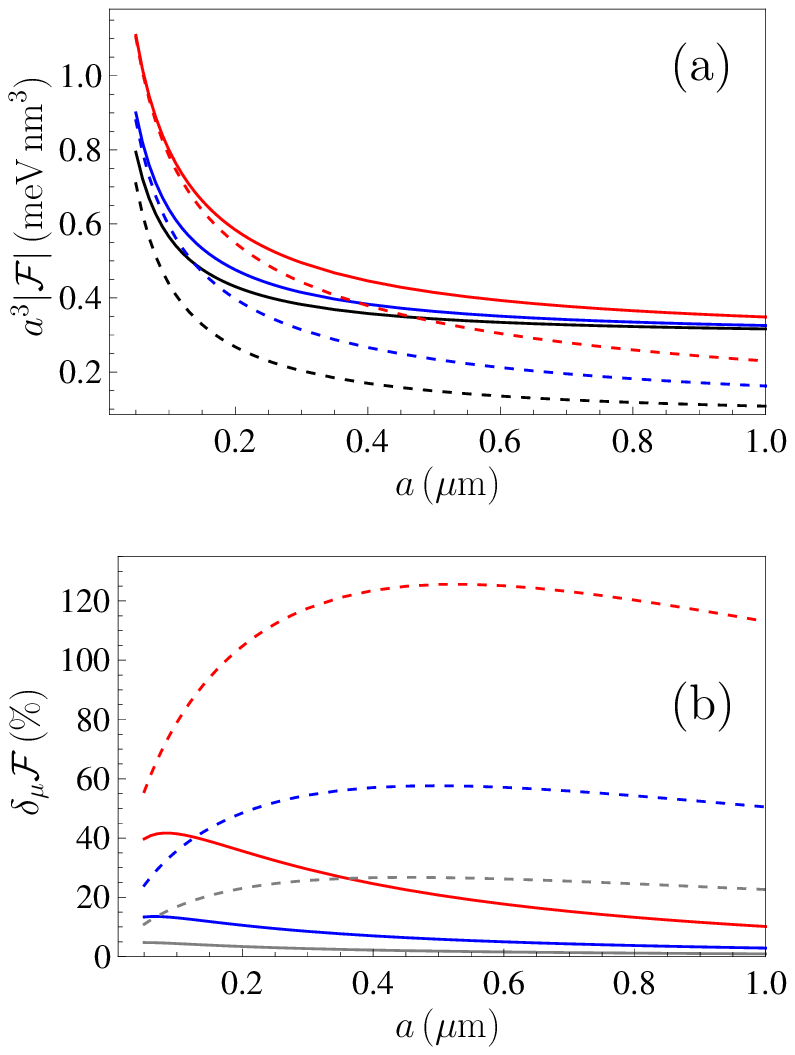}
}
\vspace*{-16.5cm}
\caption{\label{fg2}
(a) The magnitudes of the Casimir-Polder free energy multiplied by
the third power of separation between an atom of \hem and a gapless
graphene are shown as functions of separation at $T=300\,$K
(the solid lines) and  $T=77\,$K (the dashed lines). The lines of
each kind from bottom to top are plotted for graphene with $\mu=0$,
0.2, and 0.5\,eV, respectively.
(b) The relative changes in the Casimir-Polder free energy due to
nonzero chemical potential are shown as functions of separation
by the solid and dashed lines at $T=300\,$K and  $T=77\,$K,
respectively. The lines of
each kind from bottom to top are  for graphene with $\mu=0.1$,
0.2, and 0.5\,eV, respectively.
}
\end{figure}

In Fig.~\ref{fg2}(a) we present the computational results for the magnitude of
the Casimir-Polder free energy as functions of separation.
For the ease of representing
the data, we multiply them with the third power of separation between
an atom of \hem and a graphene sheet.
The computations are done for a gapless graphene ($\Delta=0$) at room temperature
$T=300\,$K (the solid lines) and at liquid nitrogen temperature $T=77\,$K
(the dashed lines).
We use a value for the Fermi velocity of
$v_F \approx c/300$.
For the three solid (and three dashed) lines considered
from bottom to top the chemical potential of graphene $\mu$ is equal to
0, 0.2, and 0.5\,eV, respectively.

Note that the chemical potential can be expressed via the doping concentration $n$ \cite{58a}
\begin{equation}
\mu=\hbar v_F\sqrt{\pi n}.
\label{eq18}
\end{equation}
\noindent
In Ref.~\cite{34}, $n$ was
estimated as $\approx 1.2\times 10^{10}\,\mbox{cm}^{-2}$
for nearly undoped graphene under high vacuum conditions.
This leads to a chemical potential that does not exceed a
value of $\mu \approx 0.02\,$eV. The values of $\mu=0.2$ and 0.5\,eV occur
for doping concentrations
$n\approx 3\times 10^{12}$ and $2\times 10^{13}\,\mbox{cm}^{-2}$.

As is seen in Fig.~\ref{fg2}(a), the magnitude of the Casimir-Polder free
energy quickly decreases with increasing atom-graphene separation. In so doing at all considered separations and temperatures the magnitude of the free energy is larger for
graphene with larger  chemical potential.
This result is in agreement with the result of Ref.~\cite{52} obtained for the Casimir interaction of
a graphene sheet with an ideal-metal plane.
{}From Fig.~\ref{fg2}(a) it is also seen that the distance between the pair of
bottom solid and dashed lines is much larger than between the pair of top ones.
This mean that the thermal contribution to the Casimir-Polder interaction decreases
with increasing chemical potential. Note also that it becomes
significant at much shorter distances compared to the Casimir-Polder
interaction with a bulk substrate.

In Fig.~\ref{fg2}(b) we especially consider the relative change in the
Casimir-Polder free energy which occurs when the chemical potential of a
graphene sheet becomes not equal to zero
\begin{equation}
\delta_{\mu}{\cal F}(a,T)=
\frac{{\cal F}(a,T,\mu)-{\cal F}(a,T,0)}{{\cal F}(a,T,0)}.
\label{eq19}
\end{equation}
\noindent
The quantity $\delta_{\mu}{\cal F}(a,T)$, as a function of separation,
is plotted in Fig.~\ref{fg2}(b)  at $T=300\,$K (the solid lines) and
$T=77\,$K (the dashed lines).
The solid (and dashed) lines considered from bottom to top correspond
to $\mu =0.1$, 0.2, and 0.5\,eV, respectively.
{}From Fig.~\ref{fg2}(b) it can be seen that at lower temperature
($T=77\,$K) the relative differences in the Casimir-Polder free energy
are much larger than at $T=300\,$K. It is also seen that at $T=77\,$K
all the dashed lines reach the maximum value at some separation,
whereas at $T=300\,$K only the top solid line reaches the maximum value in
the considered separation region from 50\,nm to $1\,\mu$m.

\begin{figure}[b]
\vspace*{0cm}
\centerline{\hspace*{2.5cm}
\includegraphics{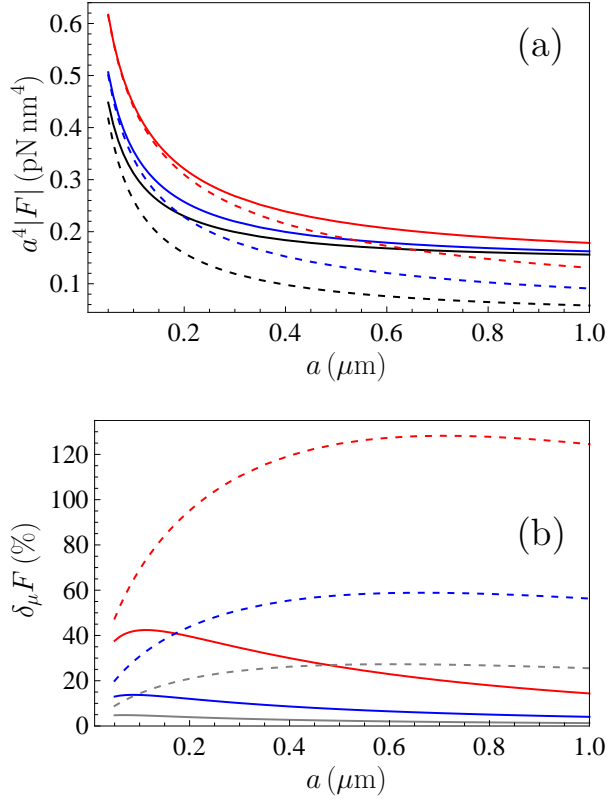}
}
\vspace*{-16.5cm}
\caption{\label{fg3}
(a) The magnitudes of the Casimir-Polder force multiplied by
the fourth power of separation between an atom of \hem and a gapless
graphene are shown as functions of separation at $T=300\,$K
(the solid lines) and  $T=77\,$K (the dashed lines). The lines of
each kind from bottom to top are plotted for graphene with $\mu=0$,
0.2, and 0.5\,eV, respectively.
(b) The relative changes in the Casimir-Polder force due to
nonzero chemical potential are shown as functions of separation
by the solid and dashed lines at $T=300\,$K and  $T=77\,$K,
respectively. The lines of
each kind from bottom to top are  for graphene with $\mu=0.1$,
0.2, and 0.5\,eV, respectively.
}
\end{figure}

Similar results for the Casimir-Polder force computed using Eq.~(\ref{eq16})
are presented in Fig.~\ref{fg3} using the same parameters and the
same notation for all lines.
As is seen in Fig.~\ref{fg3}(a), the magnitude of the Casimir-Polder force
decreases
even more quickly than the free
energy with increasing separation. At each separation it is larger
for larger chemical potential.
As it holds for the free energy, the thermal effect in the Casimir-Polder force
is larger for graphene with lower chemical potential.

In Fig.~\ref{fg3}(b) the relative change in the
Casimir-Polder force due to nonzero chemical potential,
defined in the same way as in Eq.~(\ref{eq19}),
is plotted as a function of separation at $T=300\,$K (the solid lines) and
$T=77\,$K (the dashed lines) for $\mu=0.1$, 0.2, and 0.5\,eV, respectively,
when lines are considered from bottom to top. It is seen that, again, the
relative impact of nonzero $\mu$ on the Casimir-Polder force is greater
at lower temperature. The maximum values of $\delta_{\mu}F$ are reached at
larger separations than for  $\delta_{\mu}{\cal F}$.

Now we consider the Casimir free energy between an atom of \hem and a graphene
sheet with different values of mass gap $\Delta=2mc^2$ as a function of the
chemical potential $\mu$. In fact for a perfect (pristine) graphene the
Dirac-type electronic quasiparticles are massless, so that $m=0$, $\Delta=0$.
For real graphene samples, however, the account of interelectron interactions,
structural defects, and the presence of a substrate
may lead to some nonzero mass gap estimated as
$\Delta\lesssim 0.2\,$eV \cite{5,59,60,61}.

\begin{figure}[b]
\vspace*{0cm}
\centerline{\hspace*{2.5cm}
\includegraphics{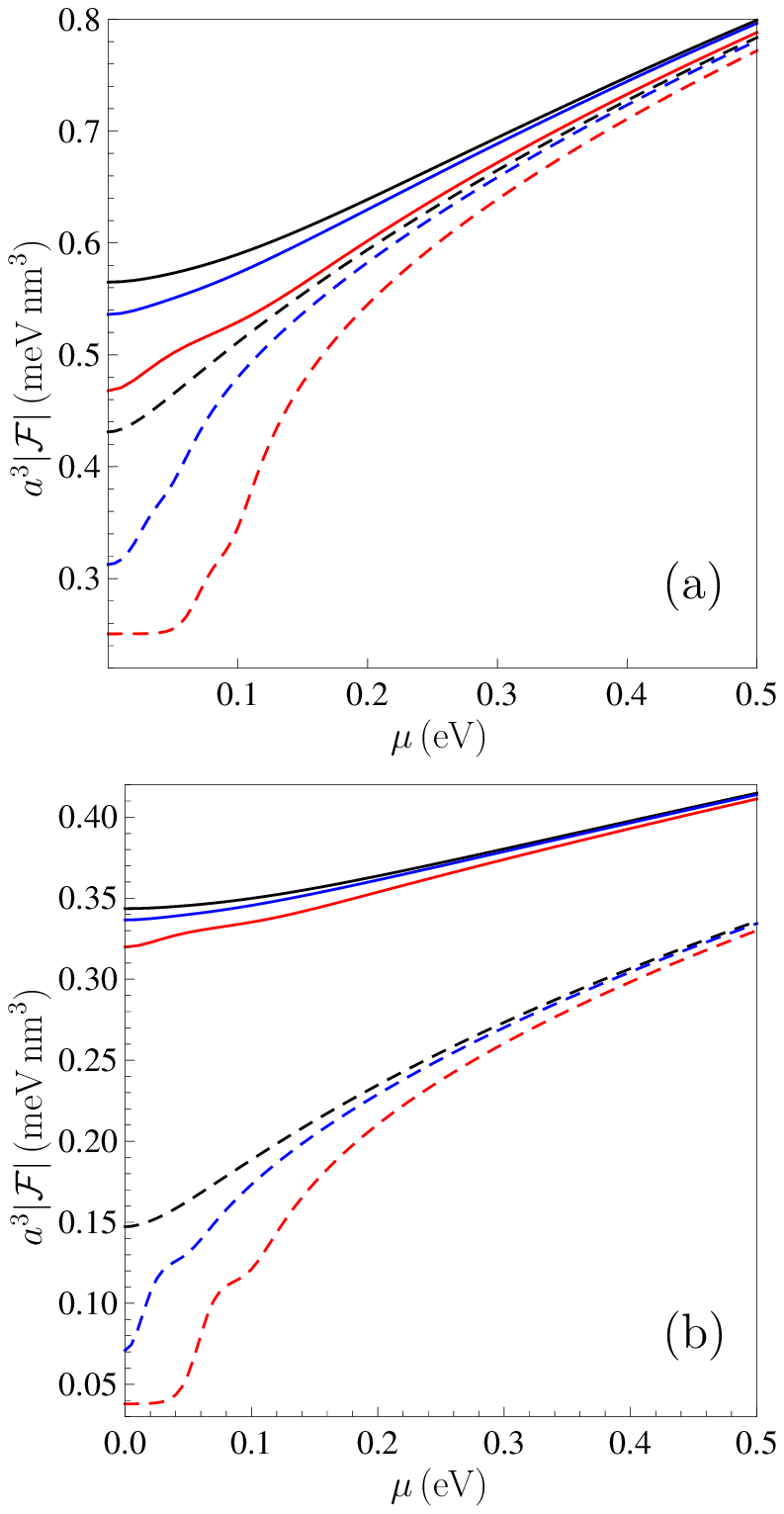}
}
\vspace*{-12.5cm}
\caption{\label{fg4}
(a) The magnitudes of the Casimir-Polder free energy multiplied by
the third power of separation between an atom of \hem and gapped
graphene are shown  at $T=300\,$K
(the solid lines) and  $T=77\,$K (the dashed lines)
as functions of chemical potential for (a) $a=100\,$nm and
(b) $a=500\,$nm. The lines of
each kind from bottom to top are plotted for graphene with $\Delta=0.2$,
0.1, and 0\,eV, respectively.
}
\end{figure}

In Fig.~\ref{fg4} we plot the magnitude of the Casimir-Polder free energy
between an atom of \hem and
a gapped graphene sheet at $T=300\,$K and
$T=77\,$K as function of the chemical potential $\mu$ at
(a) $a=100\,$nm and (b) $a=500\,$nm. For both temperatures,
the mass gap is $\Delta=0.2$, 0.1, and 0\,eV from bottom to top.
As is seen in Figs.~\ref{fg4}(a) and \ref{fg4}(b),
at all values of $\mu$ and $T$ an increase of the mass gap results in a decrease
in the magnitude of the Casimir-Polder free energy.
As to the thermal effect, it becomes larger for graphene sheets with larger
mass gap.
{}From Figs.~\ref{fg4}(a) and \ref{fg4}(b) it can be seen that the dependence of
$a^3|{\cal F}|$ on $\mu$ exhibits some kind of steps which are better seen in the
case of lower temperature (the dashed lines). This is because with decreasing $T$
the contribution of the first fraction in the first line on the right-hand side of
Eq.~(\ref{eq11}) becomes negligibly small, whereas the major part in the
contribution of the second fraction appears under the condition
\begin{equation}
B_lu-\frac{\mu}{k_BT}<0.
\label{eq21}
\end{equation}
Taking into account the definition of $B_l$ in Eq.~(\ref{eq11}) and the fact that
the integration in Eq.~(\ref{eq10}) is over the interval $u\geq D_l$, we find
from Eq.~(\ref{eq21})  that with decreasing temperature the major contribution
to $\tp_{00,l}^{(1)}$ and $\tp_l^{(1)}$ appears under the condition
$2mc^2=\Delta<2\mu$. If this condition is not satisfied, with decreasing temperature
the free energy becomes almost independent on $\mu$ \cite{53}.
The latter explains the characteristic flat regions of the dashed lines in the vicinity
of zero $\mu$. As to the second steps at nonzero $\mu$, which are clearly seen on
the dashed lines in Fig.~\ref{fg4}(b), they are explained by an interplay between
the condition $\Delta<2\mu$, which influences only partially at nonzero temperature,
and the role of thermal effects determined by the effective temperature
$T_{\rm eff}=\hbar v_F/(2ak_B)$.
This takes a value $T_{\rm eff} \approx 38\,{\rm K}$ at $a \approx 100\,{\rm nm}$.

\begin{figure}[b]
\vspace*{0cm}
\centerline{\hspace*{2.5cm}
\includegraphics{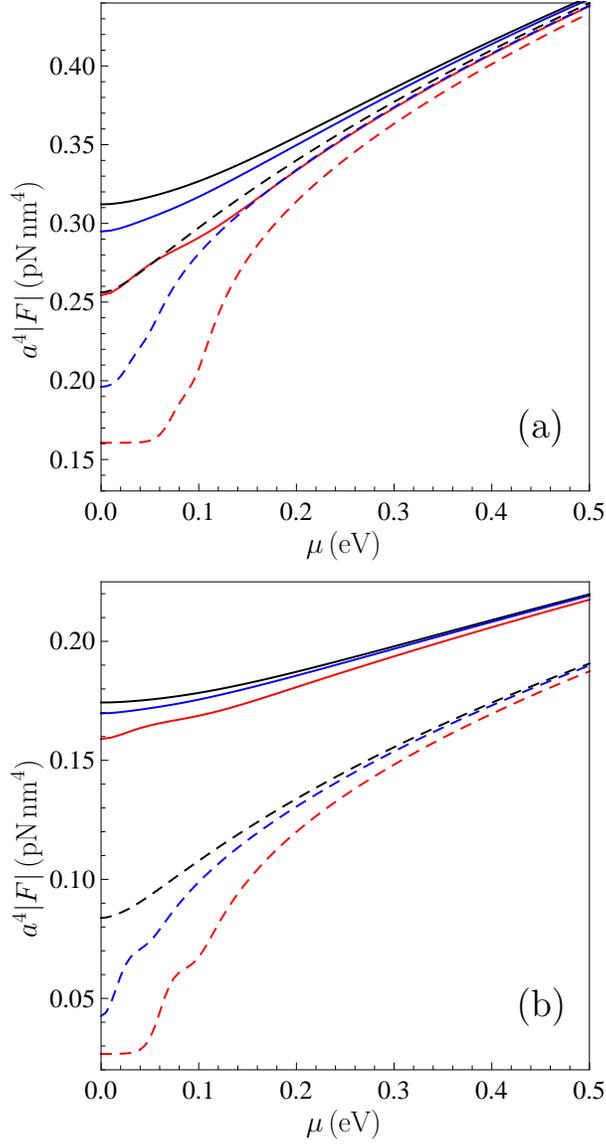}
}
\vspace*{-12.5cm}
\caption{\label{fg5}
(a) The magnitudes of the Casimir-Polder force multiplied by
the fourth power of separation between an atom of \hem and gapped
graphene are shown  at $T=300\,$K
(the solid lines) and  $T=77\,$K (the dashed lines)
as functions of chemical potential for (a) $a=100\,$nm and
(b) $a=500\,$nm. The lines of
each kind from bottom to top are plotted for graphene with $\Delta=0.2$,
0.1, and 0\,eV, respectively.
}
\end{figure}

Figure~\ref{fg5} demonstrates similar computational results for the magnitude of
the Casimir-Polder force multiplied by the fourth power of separation
between an atom of \hem and a gapped graphene sheet. All the notations are the same
as in Fig.~\ref{fg4}.
{}From Figs.~\ref{fg5}(a) and \ref{fg5}(b), plotted at $a=100$ and 500\,nm,
respectively, it is seen that with increasing mass gap
the magnitude of the Casimir-Polder force at some fixed temperature and chemical
potential becomes smaller.
The thermal effect in the Casimir-Polder force becomes larger for larger
mass-gap parameter.
The step structure clearly seen on the dashed lines plotted at $T=77\,$K is
explained by the same reasons as for the Casimir-Polder free energy.
It is interesting that at $a=100\,$nm, $\mu>0.04\,$eV
the Casimir-Polder force calculated at $T=77\,$K
for graphene with $\Delta=0$ becomes larger in magnitude than
that one at $T=300\,$K
for graphene with $\Delta=0.2\,$eV [see the top dashed and the bottom solid lines in
Fig.~\ref{fg5}(a)]. This means that the effect of chemical potential on the
Casimir-Polder force can exceed the thermal effect. In a similar way,
at $a=100\,$nm, $\mu>0.16\,$eV the Casimir-Polder force  calculated at $T=77\,$K
for graphene with $\Delta=0.1\,$eV reaches and remains equal to that one
at $T=300\,$K for graphene with $\Delta=0.2\,$eV [compare the bottom solid and the
first above the bottom dashed lines in Fig.~\ref{fg5}(a)].

To conclude this section we note that the thermal dependence
of the Casimir-Polder free energy and force consists of two
contributions.
The first one originates from the summation over the Matsubara
frequencies in the Lifshitz formulas (\ref{eq1}) and (\ref{eq16})
when the polarization tensor at zero temperature is used in
computations.
The second one is from an explicit temperature dependence of the
full polarization tensor defined at nonzero temperature.
In Ref.~\cite{35} it was shown that for a gapped graphene with zero
chemical potential the explicit temperature dependence of the
polarization tensor contributes to the thermal Casimir-Polder
interaction significantly. Thus, this contribution is equal to
23\% of the total Casimir-Polder free energy for a gapless
graphene at $a=1\,\mu$m, $T=300\,$K, and it increases quickly
with increasing mass gap reaching more than 80\% of the free
energy \cite{35}. Similar situation holds for graphene with
nonzero chemical potential satisfying the condition
$2\mu<\Delta$.
Under this condition the chemical potential does not
influence the value of the polarization tensor at zero
temperature \cite{53}, so that
\begin{eqnarray}
&&
\tilde{\Pi}_{00,l}(y,0,m,\mu)=\tilde{\Pi}_{00,l}^{(0)}(y,m),
\nonumber \\
&&
\tilde{\Pi}_{l}(y,0,m,\mu)=\tilde{\Pi}_{l}^{(0)}(y,m)
\label{eq21a}
\end{eqnarray}
\noindent
for $\mu < m c^2$. In this regime, the quantities
$\tilde{\Pi}_{00,l}^{(1)}$, $\tilde{\Pi}_{l}^{(1)}$
in Eq.~(\ref{eq7}) vanish when $T\to 0$ and, thus,
coincide with
the thermal corrections.
If, however, the condition $2\mu>\Delta$ is satisfied,
the polarization tensor becomes $\mu$-dependent
even at zero temperature:
\begin{eqnarray}
&&
\tilde{\Pi}_{00,l}(y,0,m,\mu)=\tilde{\Pi}_{00,l}^{(0)}(y,m)+
\tilde{\Pi}_{00,l}^{(1)}(y,0,m,\mu),
\nonumber \\
&&
\tilde{\Pi}_{l}(y,0,m,\mu)=\tilde{\Pi}_{l}^{(0)}(y,m)+
\tilde{\Pi}_{l}^{(1)}(y,0,m,\mu).
\label{eq21b}
\end{eqnarray}
\noindent
The second terms on the right-hand side of Eq.~(\ref{eq21b})
increase the polarization tensor at $T=0$ significantly.
As a result, the explicit temperature dependence of the
polarization tensor plays a weaker role and the Casimir-Polder
free energy and force become temperature-dependent mainly
because of the Matsubara summation.
Thus, for $\Delta=0$ and $\mu=0.1\,$eV
the explicit dependence of the polarization tensor on $T$
contributes only 0.24\% and 0.12\% to the total Casimir-Polder
free energy at $a=100\,$nm and $1\,\mu$m, respectively.
If $\mu =0.2\,$eV, this contribution is equal to 0.24\%
and 0.06\% at the same respective separations.
One can conclude that with respect to the relative roles of two
thermal contributions to the Casimir-Polder free energy,
 the mass gap and chemical potential
again influence in the opposite directions.

In the next section we consider how the above results
for the Casimir-Polder free energy and force are modified if the
graphene sheet is deposited on a material substrate.

\section{Interaction with substrate coated with doped graphene}

Here, we consider the Casimir-Polder interaction of a \hem atom with a
graphene sheet
deposited on an amorphous silica (SiO${}_2$) substrate. This material is often used
for the deposition of graphene \cite{34,62}.
Computations of the Casimir-Polder free energy and force using Eqs.~(\ref{eq1}) and
(\ref{eq16}) with the reflection coefficients (\ref{eq5}) require the values of the
dielectric permittivity of SiO${}_2$, $\varepsilon_l$, at the imaginary Matsubara
frequencies. A sufficiently exact approximation for $\varepsilon_{{\rm SiO}_2}$
along the imaginary frequency axis is given by the two-oscillator model \cite{63,64}
\begin{equation}
\varepsilon_{{\rm SiO}_2}(i\xi)=1+
\frac{C_{\rm UV}\omega_{\rm UV}^2}{\xi^2+\omega_{\rm UV}^2}+
\frac{C_{\rm IR}\omega_{\rm IR}^2}{\xi^2+\omega_{\rm IR}^2},
\label{eq22}
\end{equation}
\noindent
with the ``oscillator strengths''
$C_{\rm UV}=1.098$, $C_{\rm IR}=1.703$
and ``resonance frequencies''
$\omega_{\rm UV}=2.033\times 10^{16}\,$rad/s, and
$\omega_{\rm IR}=1.88\times 10^{14}\,$rad/s.

We begin by considering the change of the
Casimir-Polder free energy when the SiO${}_2$ plate is coated with a
graphene sheet. For this purpose we calculate the dimensionless ratio
\begin{equation}
\delta_g{\cal F}_{\!{\rm SiO}_2}(a,T)=
\frac{{\cal F}_{\!{\rm SiO}_2}^{\,g}(a,T)-
{\cal F}_{\!{\rm SiO}_2}(a,T)}{{\cal F}_{\!{\rm SiO}_2}(a,T)},
\label{eq22}
\end{equation}
\noindent
where
${\cal F}_{\!{\rm SiO}_2}^{\,g}$ and ${\cal F}_{\!{\rm SiO}_2}$
are the free energies of an atom interacting with the graphene-coated and
uncoated SiO${}_2$ plates, respectively.
The quantity ${\cal F}_{\!{\rm SiO}_2}$ is calculated using the same Lifshitz
formula (\ref{eq1}), as ${\cal F}_{\!{\rm SiO}_2}^{\,g}$, but in the reflection
coefficients (\ref{eq5}) one should put $\tp_{00,l}=\tp_l=0$ and obtain
\begin{eqnarray}
&&
R_{\rm TM}^{{\rm SiO}_2}(i\zeta_l,y)=
\frac{\ve_ly-\sqrt{y^2+(\ve_l-1)\zeta_l^2}}{\ve_ly+
\sqrt{y^2+(\ve_l-1)\zeta_l^2}},
\nonumber \\[1mm]
&&
R_{\rm TE}^{{\rm SiO}_2}(i\zeta_l,y)=
\frac{y-\sqrt{y^2+(\ve_l-1)\zeta_l^2}}{y+
\sqrt{y^2+(\ve_l-1)\zeta_l^2}}.
\label{eq24}
\end{eqnarray}
First, we focus on gapless ($\Delta = 0$) graphene
and vary the chemical potential.

\begin{figure}[b]
\vspace*{1cm}
\centerline{\hspace*{2.5cm}
\includegraphics{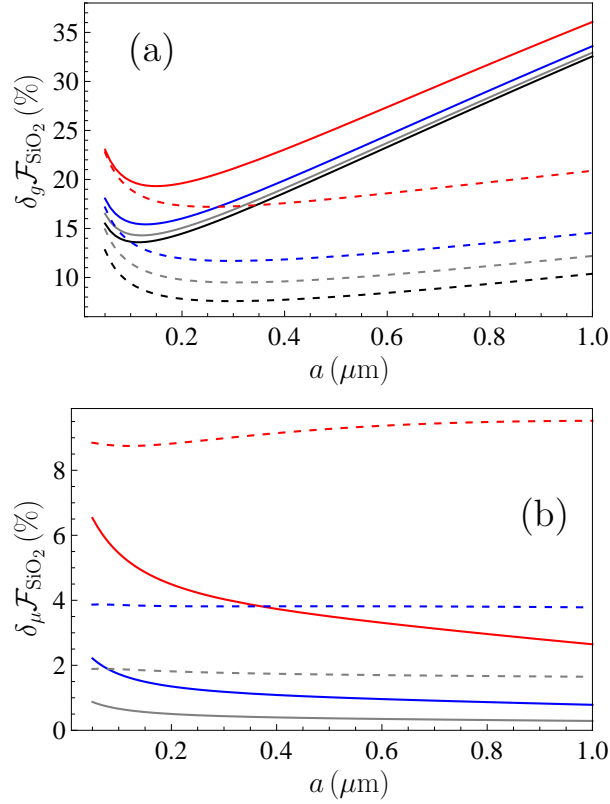}
}
\vspace*{-16.5cm}
\caption{\label{fg6}
The relative changes in the Casimir-Polder free energy for
an atom of \hem interacting with a gapless graphene
deposited on a SiO${}_2$ substrate due to (a) the presence
of graphene (the solid and dashed lines considered from
bottom to top are for $\mu$=0, 0.1, 0.2, and 0.5 eV,
respectively) and (b) the nonzero chemical potential (the
solid and dashed lines considered from bottom to top are
for $\mu$=0.1, 0.2, and 0.5 eV, respectively) are shown
as functions of separation. The solid
lines are computed at $T=300\,$K and the dashed ones at
$T=77\,$K.
}
\end{figure}

In Fig.~\ref{fg6}(a), the computational results for the relative change
$\delta_g{\cal F}_{\!{\rm SiO}_2}$ due to the presence of graphene are
shown as functions of separation by the solid and dashed lines at
$T=300\,$K and 77\,K, respectively.
Both the solid and dashed lines, considered from the bottom one to the
top one, correspond to the chemical potential, equal to 0, 0.1, 0.2, and
0.5\,eV, respectively. As is seen in Fig.~\ref{fg6}(a), the
Casimir--Polder free energy increases in magnitude by coating
the silica substrate with graphene.
For each value
of $\mu$ the quantity $\delta_g{\cal F}_{\!{\rm SiO}_2}$ reaches a
minimum value at rather short separations and then increases with increasing
$a$ rather quickly (at $T=300\,$K) or slowly (at $T=77\,$K).
For larger $\mu$ the relative change  $\delta_g{\cal F}_{\!{\rm SiO}_2}$
takes larger values at all separations.

In Fig.~\ref{fg6}(b) the computational results for the relative change
$\delta_\mu{\cal F}$ [Eq.~(\ref{eq19})] in the Casimir-Polder free energy, which occurs when $\mu$
becomes not equal to zero, are plotted for the case of gapless graphene
deposited on a SiO${}_2$ substrate.
Note that in the definition of this quantity  (\ref{eq19}) ${\cal F}$
on the right-hand side should be replaced with
${\cal F}_{\!{\rm SiO}_2}^{\,g}$
and $\delta_{\mu}{\cal F}$ with $\delta_{\mu}{\cal F}_{\!{\rm SiO}_2}$.
The computational results for $\delta_{\mu}{\cal F}_{\!{\rm SiO}_2}$
are shown in Fig.~\ref{fg6}(b) by the solid lines at $T=300\,$K and by
the dashed lines at $T=77\,$K as functions of separation.
The lines of both kinds counted from bottom to top correspond to
$\mu =0.1$, 0.2, and 0.5\,eV, respectively. Note that with decreasing
temperature (the dashed lines) the relative change in the Casimir-Polder
free energy due to nonzero $\mu$ for graphene deposited on a substrate
becomes almost independent on separation. This is different from the case
of a free-standing graphene sheet [see Fig.~\ref{fg2}(b)].

\begin{figure}[b]
\vspace*{0cm}
\centerline{\hspace*{2.5cm}
\includegraphics{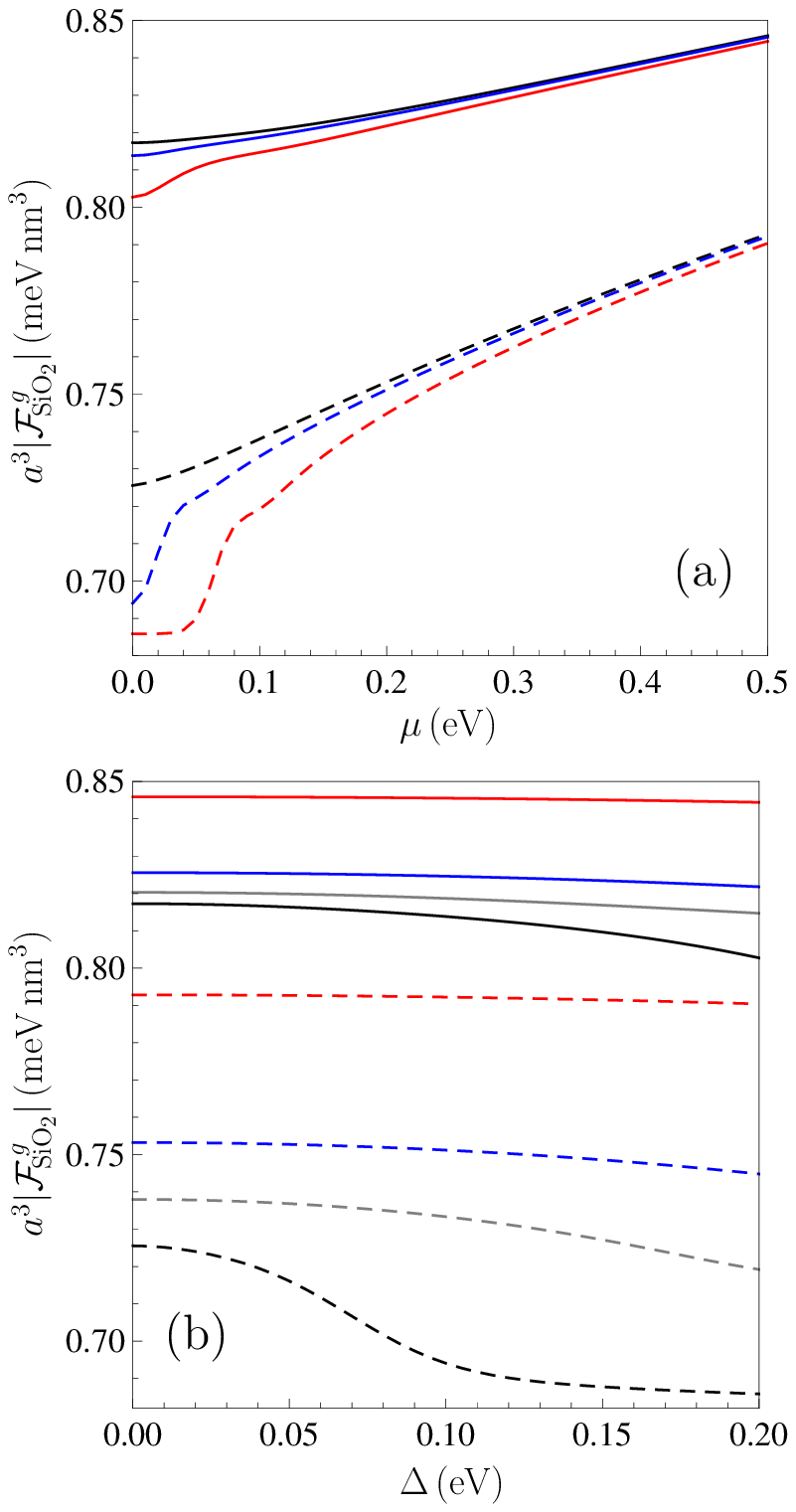}
}
\vspace*{-12.5cm}
\caption{\label{fg7}
The magnitudes of the Casimir-Polder free energy multiplied
by the third power of separation $a$=0.5$\mu$m between an atom
of \hem and graphene-coated SiO${}_2$ substrate are shown at
$T=300\,$K (the solid lines) and at $T=77\,$K (the dashed lines)
as functions of (a) the chemical potential and (b) the mass-gap
parameter. The lines of each kind from bottom to top are plotted
for graphene with (a) $\Delta=0.2$, 0.1, and 0\,eV and (b)
$\mu=0.01$, 0.2, and 0.5\,eV, respectively.
}
\end{figure}

Next we consider the dependence of the Casimir free energy on both the
chemical potential and mass-gap parameter of graphene coating.
In Fig.~\ref{fg7} the magnitudes of the free energy of an atom of \hem
interacting with a graphene-coated SiO${}_2$ substrate multiplied by
the third power of separation $a=0.5\,\mu$m are plotted by the solid
and dashed lines at $T=300\,$K and  $T=77\,$K, respectively, as functions of
(a) chemical potential $\mu$ and (b) mass-gap parameter $\Delta$.
The lines of each type counted from bottom to top
correspond to (a) $\Delta=0.2$, 0.1, and 0\,eV and
(b) $\mu=0$, 0.1, 0.2, and 0.5\,eV.

As is seen in Fig.~\ref{fg7}(a), with increasing $\Delta$ the magnitude of the
Casimir-Polder free energy decreases, and this effect is more
pronounced at lower temperature.
This is in agreement with the case of a free-standing graphene
sheet. The comparison of Fig.~\ref{fg7}(a) with Fig.~\ref{fg4}(b) plotted for
a free-standing graphene at the same distance from an atom $a=0.5\,\mu$m shows
that for a graphene-coated substrate we have a strong increase
in the magnitude of the free
energy. This is explained by the role of
SiO${}_2$ substrate. At the same time, the fine structure of the lines
(including the steps typical for the dashed lines discussed in Sec.~III) is
caused by the presence of graphene coating.

{}From Fig.~\ref{fg7}(b) one concludes that the impact of chemical potential
on the Casimir-Polder free energy is just the opposite, as compared to the
impact of $\Delta$. Specifically, with increasing $\mu$ the magnitude of the
free energy $|{\cal F}_{\!{\rm SiO}_2}^{\,g}|$ increases and the thermal
correction becomes smaller. With increasing mass-gap parameter,
$|{\cal F}_{\!{\rm SiO}_2}^{\,g}|$
decreases slowly when the chemical potential
is relatively large ($\mu =0.5$ or 0.2\,eV) and more rapidly for $\mu=0.1\,$eV
or for undoped graphene.

\begin{figure}[b]
\vspace*{0cm}
\centerline{\hspace*{2.5cm}
\includegraphics{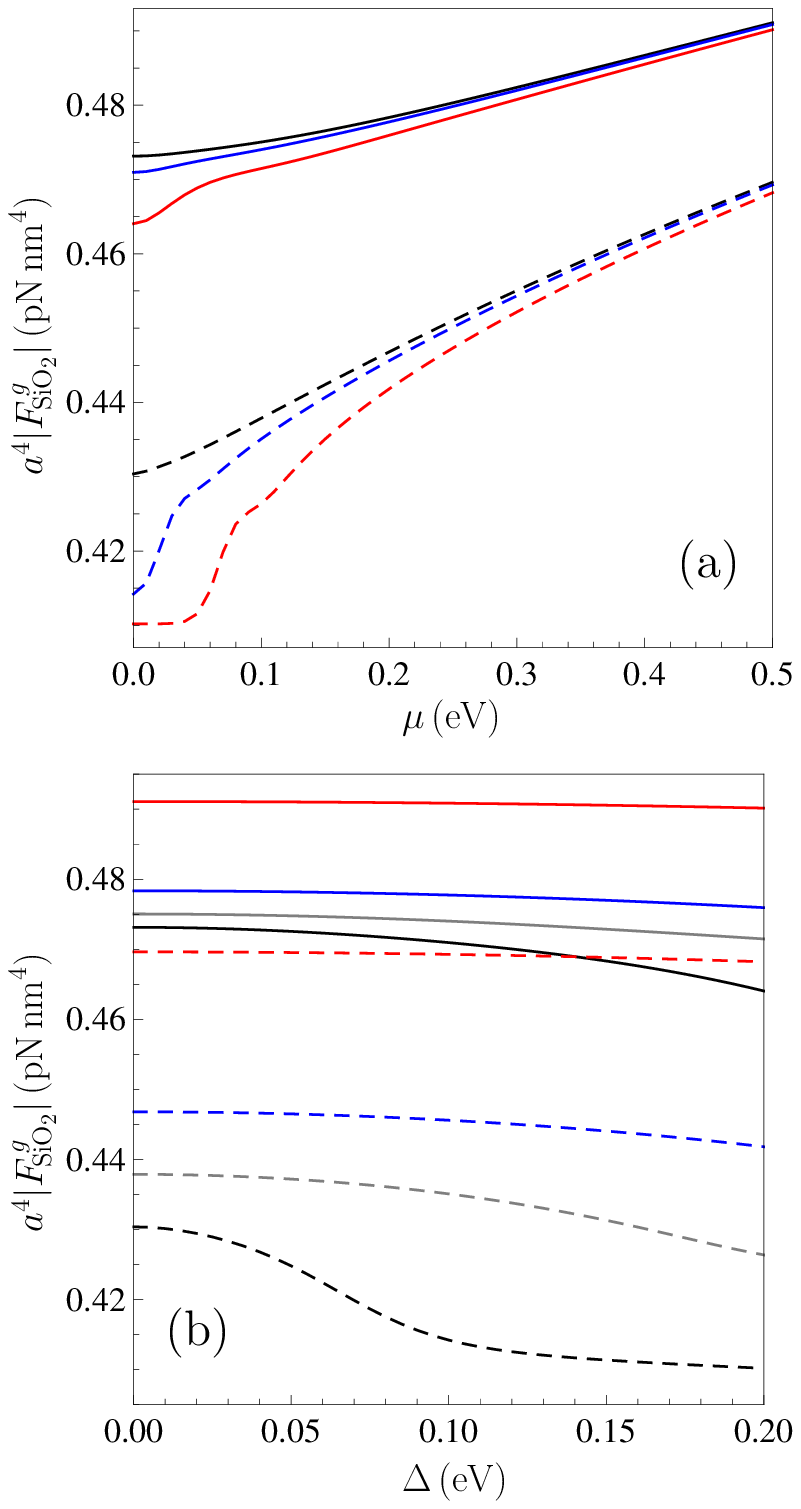}
}
\vspace*{-12.5cm}
\caption{\label{fg8}
The magnitudes of the Casimir-Polder force multiplied
by the fourth power of separation $a$=0.5$\mu$m between an atom
of \hem and graphene-coated SiO${}_2$ substrate are shown at
$T=300\,$K (the solid lines) and at $T=77\,$K (the dashed lines)
as functions of (a) the chemical potential and (b) the mass-gap
parameter. The lines of each kind from bottom to top are plotted
for graphene with (a) $\Delta=0.2$, 0.1, and 0\,eV and (b)
$\mu=0.01$, 0.2, and 0.5\,eV, respectively.
}
\end{figure}

Similar computations have been performed for the Casimir-Polder force between
an atom of \hem and
a graphene-coated SiO${}_2$ substrate.
Using the same notations, as in Fig.~\ref{fg7}, the computational results for
the magnitude of the Casimir-Polder force  multiplied by
the fourth power of separation $a=0.5\,\mu$m are shown as functions of the
chemical potential $\mu$ in Fig.~\ref{fg8}(a) and as  functions of the
mass-gap parameter in Fig.~\ref{fg8}(b).
As can be seen in Figs.~\ref{fg8}(a) and \ref{fg8}(b), qualitatively the character
of the quantity $a^4|{F}_{\!{\rm SiO}_2}^{\,g}|$ is the same as
$a^3|{\cal F}_{\!{\rm SiO}_2}^{\,g}|$.
Specifically, the increase of $\mu$ and $\Delta$ impact on
$|{F}_{\!{\rm SiO}_2}^{\,g}|$ in the opposite directions by increasing and
decreasing it, respectively. The comparison of Fig.~\ref{fg8}(a)  with
Fig.~\ref{fg5}(b) again demonstrates that the main contribution to the force
magnitude is given by the substrate, whereas the fine structure of the force
lines is caused by the graphene coating. We also note that the
lines for the Casimir-Polder force for $77\,$K and $300\,$K
cross [compare the top dashed line with the bottom solid line in Fig.~\ref{fg8}(b)].
This illustrates the wide tuning range that becomes available
by changing the chemical potential.


\section{Conclusion and discussion}

In this paper we have investigated the Casimir-Polder
interaction between an atom and a free-standing
graphene sheet characterized by some chemical
potential and mass-gap parameter. The interaction of
an atom with a graphene-coated substrate was also
considered. For this purpose, we used the exact formalism of
the polarization tensor in (2+1)-dimensional space-time,
developed in the framework of the Dirac
model. This approach permits a detailed analysis of the
impact of nonzero chemical potential of gapped graphene
on the Casimir-Polder free energy and force at
arbitrary temperature at not too short separations
between an atom and a graphene sheet or a graphene-%
coated substrate. Keeping in mind that during the last
few years the Casimir-Polder interaction has found
numerous and diverse applications (see, e.g.,
Refs.~\cite{53b,53c,65,66,67,68,69,70,71,72,73}), the elaboration of
the corresponding theoretical formalism for graphene, started
in Refs.~\cite{19,20,21,22,23,23a,35,36,37,38,39,40},
should be considered as rather promising.

The developed theory was applied to compute the
Casimir-Polder free energy and force between an atom
of metastable helium \hem and graphene sheets
with various values of the chemical potential $\mu$
and mass-gap parameter $\Delta$. 
These computations have been made possible by the use of 
Eq.~(\ref{eq6}) combined with Eqs.~(\ref{eq12}) and (\ref{eq14})
accounting for the mass gap and chemical potential.
It was shown that
with increasing $\mu$ the magnitudes of both the
Casimir-Polder free energy and force increase. By
contrast, with increasing $\Delta$ the magnitudes of
both the free energy and force decrease. Thus, the
impacts of nonzero $\mu$ and $\Delta$ on the Casimir-%
Polder interaction in real graphene samples partially
compensate each other. It was also shown that for
graphene with larger $\mu$ the thermal effect in the
Casimir-Polder interaction is smaller, whereas for
graphene with larger mass gap $\Delta$ the thermal
effect is larger.

The obtained results can be
understood qualitatively on simple physical grounds.
The point is that by increasing $\mu$ the size of the Fermi 
surface grows, thus  increasing the density of charge carriers 
and  the electrical conductivity of graphene. It is then 
quite natural that the reflection amplitudes and 
the Casimir-Polder force
increase in magnitude. By contrast, an increase
of the mass gap decreases the mobility of charge
carriers, which, in turn, decreases the conductivity
and, thus, the force magnitude. We have also found
that with decreasing temperature the functional
dependences of both the Casimir-Polder free energy
and force possess some kind of step structure
depending on the relationship between the values of
$\Delta$ and 2$\mu$.

Similar computations of the Casimir-Polder free
energy and force have been performed for an atom
of \hem interacting with a graphene sheet
deposited on a SiO${}_2$ substrate. 
These computations have been made possible by the use of
Eq.~(\ref{eq5}) combined with Eqs.~(\ref{eq12}) and (\ref{eq14}).
Qualitatively
the same results, as for a free-standing graphene
sheet, were obtained with the only difference that
the magnitudes of both the Casimir-Polder free
energy and force are much larger due to the role
of a substrate. Specifically, it was shown that
the nonzero chemical potential and mass gap act on
the Casimir-Polder interaction in the opposite
directions and partially compensate each other.

The obtained results allow a reliable calculation
of the Casimir-Polder interaction between any atom
and real graphene sheets, both free-standing and
deposited on substrates made of different materials.
These results demonstrate the possibility of tuning
the Casimir--Polder interaction over a relatively wide range
by changing the doping concentration in graphene.
This may be useful for future experiments probing
the interaction of atoms with graphene and other
two-dimensional nanostructures.

\section*{Acknowledgments}

C.H. acknowledges support from the \emph{Deutsche
Forschungsgemeinschaft} through the DIP program
(grant number Schm-1049/7-1).
G.L.K. and V.M.M. thank the University of Potsdam, where this
work was completed, for kind hospitality and partial support.
The work of V.M.M. was partially supported by the Russian
Government
Program of Competitive Growth of Kazan Federal University.


\newpage

\end{document}